\documentclass[11pt]{article}
\usepackage{fleqn,cospar}

\usepackage{url}

\usepackage{graphicx}
\usepackage[figuresright]{rotating}

\title{X-RAY SPECTROSCOPY OF CATACLYSMIC VARIABLES}

\author{Koji Mukai\address{Code 662, NASA/Goddard Space Flight Center,
      Greenbelt, MD 20771, USA; also Universities Space Research Association}}

\begin{document}

\maketitle

\begin{abstract}
In cataclysmic variables (CVs), accretion onto white dwarfs produces
high temperature, high density plasmas.  They cool down from kT$\sim$10 keV
via bremsstrahlung continuum and K and L shell line emissions.
The small volume around white dwarfs means that the plasma densities are
much higher than in, e.g., stellar coronae, probably beyond the range
well-described by existing models.  I will describe potential diagnostics
of the temperatures, the densities, and the optical depths of X-ray emitting
plasmas in CVs, and present the recent Chandra grating spectra of the magnetic
CV V1223 Sgr as an example.
\end{abstract}

\section*{MOTIVATION}

Cataclysmic variables (CVs) are interacting binaries in which the
accreting object (the primary) is a white dwarf (see Warner 1995
for a comprehensive review), unlike in X-ray binaries in which the
accreting star is a neutron star or a black hole.  CVs are numerous,
as the following line of reasoning illustrates.  In the Galaxy, there
are about 35 classical novae per year (Shafter 1997).  Since classical
nova eruptions require the primary to have accreted
$\sim 1 \times 10^{-4}$ M$_\odot$ of hydrogen-rich fuel, we infer
the total accretion rate in CVs throughout our Galaxy to be about
3.5$\times 10^{-3}$ M$_\odot$\,yr$^{-1}$.  If the mean accretion rate
per CV is 3.5$\times 10^{-10}$ M$_\odot$\,yr$^{-1}$,
then the total number of CVs in our Galaxy is about 10 million.
Although such calculations are fraught with uncertainties, particularly
in the mean accretion rate, and hence the total Galactic population
may differ from this estimate by a factor of 10, there is no doubt that
CVs are commonplace in our Galaxy.  If we further assume a mean X-ray
luminosity of CVs to be about 4$\times 10^{31}$ ergs\,s$^{-1}$
(Mukai \& Shiokawa 1993), the Galactic CVs might contribute
$\sim 4 \times 10^{38}$ ergs\,s$^{-1}$ of unresolved X-ray emission.

Moreover, the relatively high space density also means that the
nearest members of the class are less than 100 pc from the Earth,
close enough for detailed studies of the component stars and
the accretion processes in many wavelength ranges: thus the
CVs have proved to be an excellent laboratory for the study
of accretion processes.   Moreover, the relative shallowness
of the potential well of a white dwarf means that the accretion
disks of CVs are too cool to emit X-rays.  Rather, CVs generate
X-ray photons when material accretes onto the white dwarf.
Because of the modest luminosities, these X-rays do not
strongly influence the appearances of the accretion disks or
alter the secular evolution of the systems.  In many instances,
therefore, we are able to study the accretion disk in the optical
and in the UV, and the accretion process from the disk to the
white dwarf in the X-rays, independent of each other.  Such
decoupling in CVs can be used to our advantage in understanding these
systems, which can then be applied to a wider range of
accretion powered X-ray sources.

\section*{CV SUBCLASSES}

Historically, optical astronomers have used the eruptive
properties of CVs to classify them into dwarf
novae, old novae, and nova-like systems.  These are based on
whether they have been seen to exhibit either of the two distinct
types of outbursts known in CVs: dwarf nova outbursts, which
are sudden brightenings of the accretion disk, and classical nova
eruptions, which are thermonuclear runaways on the surface of the
white dwarf.  It is believed that dwarf nova outbursts are caused
by an instability of the disk at a low accretion rate; systems
that accrete at a high rate, and systems without disks, do not show
dwarf nova outbursts.  It is likely that all CVs undergo classical nova
eruptions once a sufficient amount of fresh fuel is accreted.  This
results in a typical recurrence time of $>$10,000 years.  Because
high accretion rate systems with massive white dwarfs have the
most frequent runaways, they therefore comprise the majority of old novae
(=systems seen to undergo a classical nova eruption).

The X-ray properties of a CV is strongly influenced by the magnetic
field of the primary, leading to an orthogonal scheme of CV
subclassification.  In Polars, or AM Her type systems, the
primary is so magnetic (10 to $>$200 MG) that no accretion disk can
form, optical/infrared cyclotron emission is strongly detected,
and the white dwarf spins synchronously with the binary orbit.
The intermediate polars (IPs, or DQ Her type systems)
are generally less magnetic, partial accretion disks are probably
present, cyclotron emission is weak or undetected, and the white
dwarf spin period is shorter than the orbital period.
In both these subclasses, accretion proceeds quasi-radially
along field lines onto the magnetic polar regions, and forms
accretion ``columns'' or ``curtains.''  There
a strong shock forms, heating the plasma to X-ray temperatures
(kT$>$ 10 keV).  The white dwarf surface below the shock is
heated to kT$\sim$20 eV by these X-rays and by direct injection
of kinetic energy by dense blobs.  The IPs are the strongest source
of 2--10 keV X-rays among CVs, with estimated luminosities in excess of
10$^{33}$ ergs\,s$^{-1}$, while the Polars are generally strong
($\sim 10^{32}$ ergs\,s$^{-1}$) soft X-ray sources.

Although less conspicuous to X-ray observers, the majority of
CVs are the non-magnetic, low accretion rate dwarf novae.
Dwarf novae in quiescence are relatively weak (a few times
$\sim 10^{31}$ ergs\,s$^{-1}$) sources of medium energy X-rays.
These X-rays are generated in a compact equatorial boundary layer
between the disk and the white dwarf surface (see, e.g.,
Patterson and Raymond 1985a; Mukai et al 1997).  There is no
doubt that the material in the
accretion disk just beyond the boundary layer has sufficient
kinetic energy stored in its Keplerian motion to produce these X-rays.
However, it is not clear exactly how the gas is heated to X-ray
emitting temperatures.  The very fact
that medium energy X-rays are observed require that the emitting
region is optically thin and the shocks are sufficiently strong
(Pringle \& Savonije 1979).  Numerical solutions obtained by
Narayan \& Popham (1993) are consistent with such a picture;
however, such studies to date suffer from a limited dimensionality
and a dependence on the unknown viscosity parameter, $\alpha$.
At higher accretion rates (dwarf novae in outburst, nova-like systems,
and old novae), the boundary layer should become optically thick
and radiate predominantly in the soft X-rays/EUV range
(Patterson and Raymond 1985b).  Although some systems do show
the expected soft component, observations show that the
high accretion rate systems also emit medium energy X-rays, perhaps
from the optically thin surface of the boundary layer.

The magnetic CVs are better studied in X-rays than the non-magnetic
systems, because they are brighter, and because the theory is more
straightforward.  I, too, will concentrate mostly on magnetic
CVs in the rest of this paper, with occasional references to non-magnetic
systems.

\section*{X-RAY EMISSION FROM MAGNETIC CVS}

\subsection*{The Aizu Model}

\begin{figure}
\includegraphics[width=110mm]{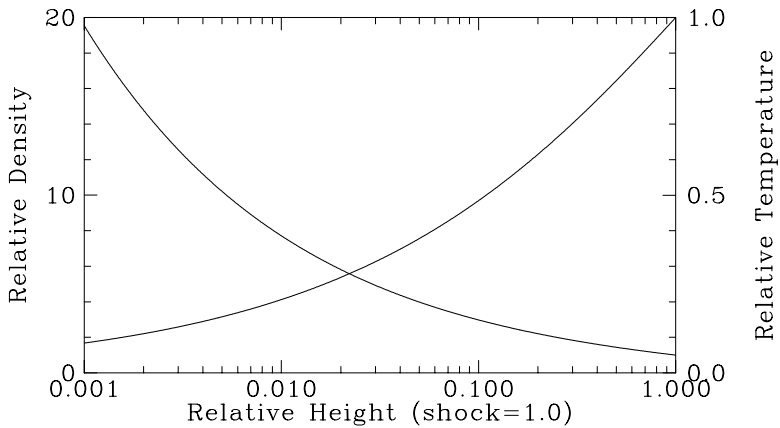}
\caption{The temperature and density profile in an Aizu shock,
as function of height above the white dwarf surface.  The relative
temperature rises to the right, density to the left.}
\end{figure}

The Aizu model was originally developed for non-magnetic white
dwarfs with a spherically symmetric accretion flow (Aizu 1973).
However, its current applications are for magnetic CVs, which
locally resemble the spherical case.  In Aizu's original paper,
numerical solutions are given in the limit of small shock height,
$h$.  There are more modern treatments, either with different
approximations, or with refinements: of the latter, the addition
of a second cooling function (Wu et al 1994) representing cyclotron
cooling is the most important.  I will use the term ``Aizu model''
inclusively, including all these variations.

In the Aizu model for magnetic CVs, material follows the field
lines to the surface, forming a strong shock.  If the shock
height is small, the shock temperature $kT_s$ is determined by
the white dwarf mass $M_{wd}$.
For accreting material of solar abundances,
$kT_s \sim$21 keV for $M_{wd}=0.6M_\odot$ and
$kT_s \sim$57 keV for $M_{wd}=1.0M_\odot$.
The shock height in turn is determined by equating the settling time
of the post-shock flow with its cooling time.  Aizu assumed optically
thin thermal bremsstrahlung to be the sole mechanism for cooling.
Although Compton cooling probably is not relevant to most magnetic CVs
(Frank et al 1983), we will return to the subject of an additional
cooling mechanism in the next section.

In the Aizu model, the emergent X-ray spectrum is the sum of emissions
from plasmas at different temperatures, from $kT_s$ to the white dwarf
photospheric temperature.  As shown in Fig.1, the bulk of the post-shock
region (note the logarithmic scale of the X axis) is hot and low-density.
At an accretion rate per unit area of 1 g\,cm$^{-2}$s$^{-1}$
onto a 0.6$M_\odot$ primary, the shock height is about 0.053
white dwarf radii (R$_{wd}$, or 460 km), and the density just below
the shock is about $1.0\times 10^{16}$ cm$^{-3}$.  Thus, it is
reasonable to ignore optical depth effects in modelling the X-ray
continuum emission from just below the shock, as long as the
accretion column radius is no more than, say, 500 km.  Although
the local accretion rate and the accretion column radius are both
poorly known, these numbers can be considered typical:  They result in
a reasonable total luminosity, which is somewhat better constrained
to be typically a few times 10$^{32}$ ergs\,s$^{-1}$ for Polars and
a few times 10$^{33}$ ergs\,s$^{-1}$ for IPs.

\subsection*{Simple X-ray spectral models}

For low signal-to-noise, low spectral resolution spectra obtained
with early X-ray satellite, however, a single temperature Bremsstrahlung
model with a simple absorber was sufficient to fit the data.  This
was mostly the case in the 1970s, except that the need to include a
reflection component was already seen in the HEAO-1 data on
the prototype Polar, AM Her (Rothschild et al 1981).  In magnetic
CVs, a solid angle of $\sim 2\pi$ steradians is subtended by the white
dwarf surface, as seen from the post-shock plasma.  Therefore, roughly
half the original X-ray photons strike the surface, resulting in
electron scattering, photoelectric absorption, and fluorescent line
emission whose relative efficiencies are a strong function of the
photon energy.  This causes a characteristic bump in the continuum
above 10 keV, as seen with HEAO-1, and the Fe K$\alpha$ line at 6.4 keV.
Indeed, emission line features in the 6--7 keV range are ubiquitous
among magnetic CVs.  However, in the proportional counter data,
it was difficult to separate the contributions of the 6.4 keV
reflection feature from those of thermal plasma at 6.7 keV (He-like Fe)
and at 6.97 keV (H-like Fe).

Norton and Watson (1989) have presented their study of
the spin modulation of IPs using EXOSAT data.
They find that the spin modulation is stronger
at lower energies, generally validating the ``accretion curtain''
model, in which the spin modulation is caused by photoelectric
absorption in the accretion flow towards the magnetic poles.
However, they also find that the energy dependence was not as
strong as would be expected from pure photoelectric absorption.
This suggests partial covering absorber: unfortunately, the
statistics of the spin phase resolved spectra of magnetic CVs
are usually not good enough to quantify this in detail, and
one cannot distinguish spatial and temporal partial covering
when analyzing phase-averaged spectra.

The Ginga LAC provided good data on magnetic CVs, often extending
up to 30 keV.  Using these, Ishida (1991) found typical Bremsstrahlung
temperatures of $\sim$15 keV for Polars and $\sim$30 keV for IPs.

\subsection*{Thermal lines from the post-shock plasma}

\begin{figure}
\includegraphics[width=110mm]{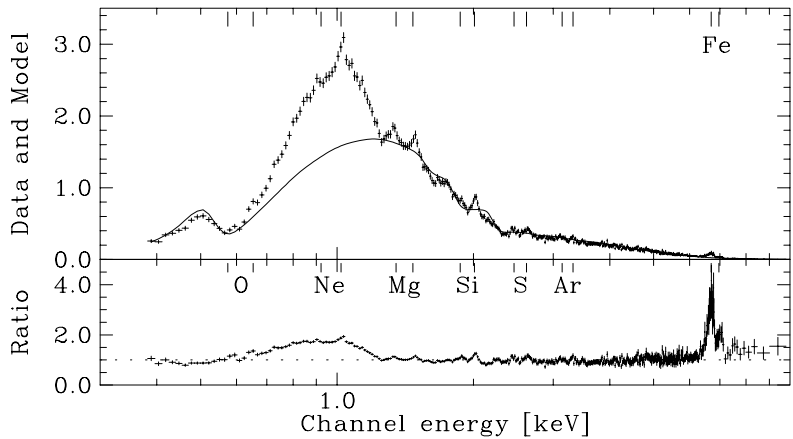}
\caption{The average ASCA SIS spectrum of the unusual IP, EX~Hya.
Thermal Bremsstrahlung plus absorber fit was performed to the data
while ignoring the 0.7--1.2 keV region.  The top panel shows
the data with this best-fit Bremsstrahlung model; the bottom
panel shows the ratio of data to the model.  Energies of H-like and
He-like K$\alpha$ lines of O, Ne, Mg, Si, S, Ar and Fe are shown.
The spectrum is clearly dominated by the Fe L complex around 1 keV.}
\end{figure}

Ishida et al (1994) observed the unusual IP, EX Hya, with the ASCA SIS.
In addition to the Fe K$\alpha$ lines, they discovered K$\alpha$ lines
of Ne, Mg, Si, S, and Ar for the first time, directly establishing the
presence of multi-temperature plasma in this system (see Fig.2).  Moreover,
something
about the temperature distribution can be inferred from the line ratios:
in fact, Fujimoto and Ishida (1997) have converted these into the shock
temperature, and estimated the primary of EX Hya to be 0.48 M$_\odot$.
They also infer the temperature at ``the bottom of the shock''
to be $\sim$0.65 keV, yet do not discuss the fact that the true
photospheric temperature must be much lower in this system
(the EUVE spectrum extends to $\sim$180\AA, without showing any
signs of the soft, blackbody-like component; see Hurwitz et al 1997).
The Fe K$\alpha$ lines in EX~Hya are resolved into
3 components, although the 6.4 keV line is rather weak in this system.
Finally, the spectrum of EX~Hya around 1 keV is dominated by the
Fe~L features, which is a characteristic of thermal plasma with
kT$\sim$1 keV.

The strength of the Fe L complex in EX~Hya (also seen in the X-ray
spectrum of stellar coronae) points out an inadequacy of the Aizu
model: the assumption of the Bremsstrahlung cooling breaks down
for optically thin, kT$\sim$1 keV plasma.  Such a plasma cools,
instead, predominantly by emitting Fe L photons (see, e.g.,
Gehrels and Williams 1993).  Thus, the Aizu model overestimates
the cooling time appropriate for the kT$\sim$1 keV region, and
overestimates its height.  More importantly, integrating plasma
emission models over the emission measure distribution calculated
using the Aizu model would result in a significant overprediction
of the Fe L feature in magnetic CVs.

However, this turns out to be the least of problems, when applied
to most magnetic CVs other than EX~Hya.  Among ASCA archival spectra
of magnetic CVs, strong Fe L features are the exception,
not the rule.  The case for this is particularly strong for the
typical IP, V1223~Sgr (Fig.3).  Even though the continuum is
strongly detected down to $\sim$0.4 keV, there is no sign of the
Fe~L features.  This, therefore, cannot be explained as due to
photoelectric absorption (which would remove continuum
and line photons equally; any photoelectric absorption which
removes the Fe~L photons will not let $\sim$0.4 keV continuum
photons through), and points to a real problem in applying the standard
spectral model to real magnetic CVs.  Of the 20 or so magnetic CVs observed
with ASCA, only EX~Hya, YY~Dra, V884~Her, and AE~Aqr show strong
Fe~L features.  Similarly, many magnetic CVs show only weak
evidence for K$\alpha$ lines of elements other than Fe.  Several
lines of arguments suggest that EX~Hya has an exceptionally low
accretion rate for an IP, and hence a relatively tall, low density
shock; this may be somehow connected to the ease with which low
energy lines have been detected in this system.

\begin{figure}
\includegraphics[width=110mm]{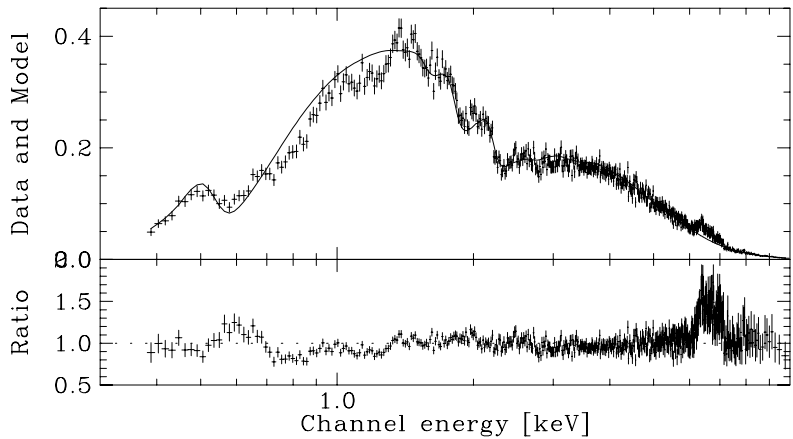}
\caption{Same as Fig.2, but for V1223 Sgr.  A strong Fe L
bump seen in EX~Hya is absent in this system.}
\end{figure}

A similar situation exists for non-magnetic CVs.  They, too, should
exhibit multi-temperature plasma emission.  There are several cases
where strong Fe K and L features are seen (e.g., V603~Aql), which is
a direct evidence for at least two temperature components; there are other
cases where the Fe~L features are absent (e.g., HT~Cas).  For both
non-magnetic and magnetic CVs, it is highly unlikely that an X-ray
emitting plasma can cool to white dwarf photospheric temperatures
without emitting Fe~L photons.  Rather, we may try to modify the
standard model by considering a geometry in which the lower temperature
component is selectively hidden, while allowing the higher temperature
component (which also emits low energy continuum photons) to escape.

The study of Fe K$\alpha$ lines have turned out to be more fruitful:
the fluorescent component can often be separated from the thermal
components, and evidence has been found for a significant broadening
of the thermal lines in some systems, while they appear narrow in
others (Hellier et al 1998).  For example, the 6.4 keV line in the IP
AO~Psc appears narrow, while the thermal lines are blended into a broad
bump; in contrast, V1223~Sgr (which is otherwise very similar to AO~Psc)
appears to have three narrow Fe K$\alpha$ lines in the ASCA data.
Having considered, and discarded, several possibilities
such as bulk Doppler motion, Hellier et al consider Compton scattering
to be the most likely broadening mechanism.  If the opacity for the
thermal line core is high enough, line photons will be trapped, and
they can only escape after Compton scattering, which degrades the
photon energy.  The possibility that resonant scattering in the post-shock
region can also enhance the equivalent width of the Fe K$\alpha$ feature,
when viewed along the magnetic axis, has been discussed by Terada et al
(1999).  The true test of this attractive idea would be a phase-resolved
observation of a system whose geometry is established from other means.
Similarly, better data as well as more detailed calculations are required
to test if the resonant trapping/Compton scattering mechanism
can quantitatively explain the broadened line profiles.

\section*{PRE-SHOCK FLOW}

If the shock height is small, then a substantial fraction of the
X-rays emitted in the post-shock region must traverse the pre-shock
flow.  

\subsection*{Absorption Effects}

As already mentioned, Norton and Watson (1989) discovered that
partial-covering absorption was necessary to model the spectrum
of IPs.  The true complexity of the intrinsic absorption is revealed
by the following exercise, divsed by Osborne (1994, private communication).
In this, one fits a simple (Bremsstrahlung plus a single, cold, absorber)
model to the ASCA spectrum of a magnetic CV, first in the entire ASCA
range ($\sim$0.4--10 keV).   Next, raise the lower limit of energy range
to 1.0, 2.0, 3.0, etc. keV and repeat the fit.  One finds that the
best fit kT and N$_H$ values change as a function of the lower limit,
revealing that the observed spectrum cannot be adequately modeled by
a simple absorber.

This is not surprising: the lines-of-sight to different parts of
the post-shock gas pass through different amounts of pre-shock flow.
One can generalize the partial covering absorber model to, say,
a power-law distribution of N$_H$ (Done and Magdziarz 1998).
Unfortunately, even this is a phenomenological model; unless we can
determine the complex geometry of the accretion flow, one cannot
create a realistic model of the complex absorption.

\begin{figure}
\includegraphics[width=135mm]{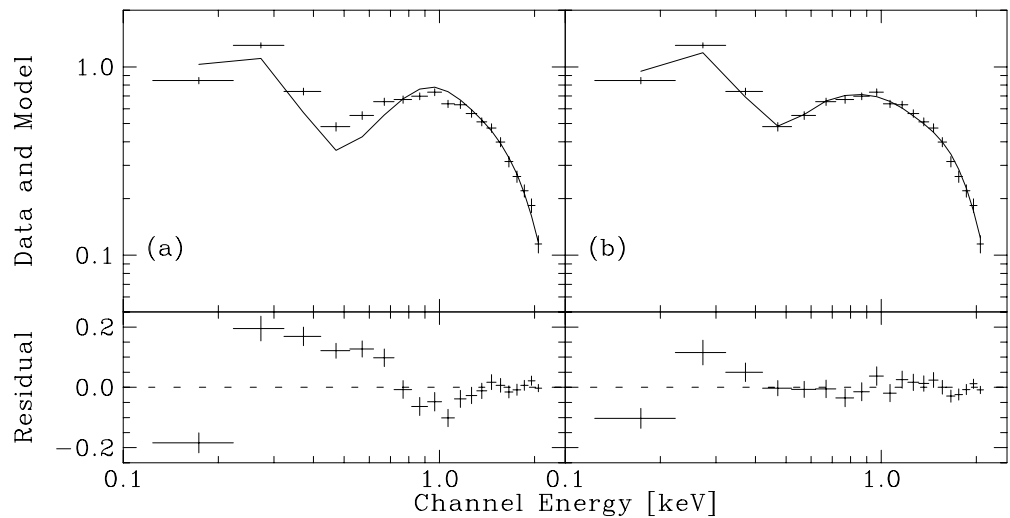}
\caption{A ROSAT PSPC spectrum of V1223 Sgr, shown twice with different
model fits.  (a) Fitted with a partial-covering absorber model;
(b) Fitted with an ionized absorber model.   The smaller residuals
for the latter suggests the presence of an ionized absorber in this
system, but the residuals can also be reduced by introducing a second
cold partial-covering absorber.}
\end{figure}

Moreover, the pre-shock flow should be ionized.  The degree of
ionization is, again, geometry-dependent: the pre-shock flow
is less ionized if more radiation can escape from the sides of
the post-shock region.  If little radiation can escape from the
sides (if accretion column radius is much larger than the
shock height, this would be the case just above the shock),
the ionization parameter $\xi$ can be several hundred.
Indeed, ASCA and ROSAT spectra of several magnetic CVs do suggest
the presence of warm absorbers, but do not uniquely prove it:
an alternative fit involving multiple partial covering absorber
can usually be found (see, e.g., Fig. 4).  To reach a final
conclusion on this issue, one would need to attempt a direct detection
of warm absorber edges in higher spectral resolution, high signal-to-noise
spectra.

\subsection*{Scattering and re-emission}

However, absorption is not the only physical process that takes
place in the pre-shock flow.  Some electron scattering will
take place, particularly if the plasma is significantly ionized;
also, photoelectric absorption will be followed by continuum and/or
line emission at lower energies.  Because the geometry of the pre-shock
flow is far from spherically symmetric around the emission region,
the relative importance of absorption vs. emission depends strongly
on the viewing geometry, which in turn depends on the binary inclination
angle and the spin phase.

In most magnetic CVs, the case for photoionized plasma emission has
been ambiguous.  For example, Kallman et al (1993) included such a
component in their interpretation of the BBXRT spectrum of BY~Cam.
However, Done and Magdziarz (1998) consider that low energy features
in this system, seen with ASCA, are due to thermal emission from
multi-temperature plasma.  The best case so far of photoionized
plasma emission may be found in the spin minimum of AM Her:
although the predominant magnetic pole is hidden behind the white
dwarf limb at this phase, a weak X-ray source remains in view
(Ishida et al 1997).  The pre-shock flow is a plausible origin
of this component.  However, emission from the second pole cannot
be excluded.

\section*{CHANDRA HETG OBSERVATION OF V1223 SGR}

In previous sections, I have reviewed the state of the art in
X-ray spectroscopy of (mostly magnetic) CVs in 1999.  Although
many recent data (notably those obtained with ASCA and BeppoSAX)
remain to be exploited fully, there are several important questions
that require higher resolution data.  Perhaps the instrument that
would have best matched the scientific objectives was the XRS
on-board ASTRO-E; nevertheless, grating instruments on XMM-Newton
and Chandra both have the potential to advance our knowledge of
CV X-ray spectra significantly.  In the rest of this article,
I will describe preliminary results from a Chandra HETG observation
of X-ray bright IP, V1223~Sgr, obtained on 2000 April 30/May 1 for
approximately 52 ksec.  Analysis and interpretation of the data
are still ongoing.

\begin{figure}
\includegraphics[width=120mm]{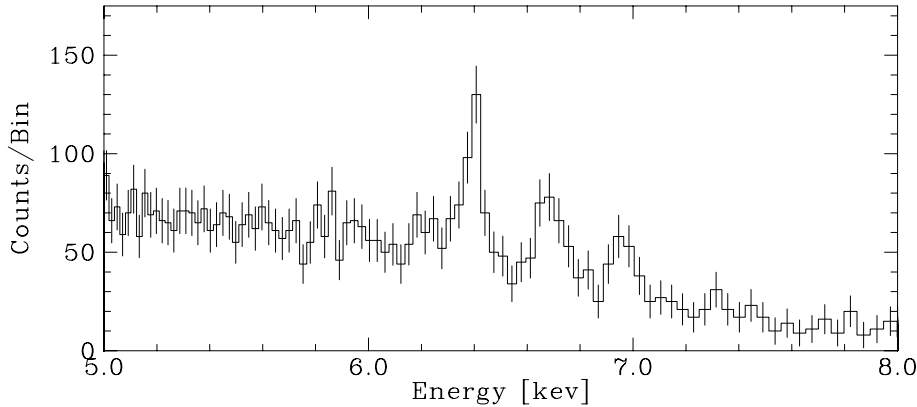}
\caption{The average Chandra HETG (HEG) spectrum of V1223 Sgr in the
5--8 keV range.  The Fe K$\alpha$ feature is clearly resolved into
three, narrow, components.  In addition, the 6.4 keV (fluorescent)
component may have a shoulder extending to $\sim$6.2 keV.}
\end{figure}

First, the Fe K$\alpha$ feature of this system clearly consists of
3 separate components (see Fig. 5).  The detected line widths
for the 6.7 and 6.97 keV thermal lines appear consistent with
instrumental resolution, confirming the ASCA result.  Thus, V1223~Sgr
should serve as a baseline; a Chandra HETG observation of AO~Psc,
the system for which the line broadening is most securely detected
in ASCA data, is scheduled in Cycle 2.

The 6.4 keV (fluorescent) component also largely appears to be narrow,
however a shoulder, extending down to $\sim$6.2 keV, is tentatively detected.
This might indicate that some 6.4 keV line photons (but not the thermal
line photons) suffer Compton degradation.  Alternatively, some of the
fluorescence line may originate in the pre-shock flow, which may have
sufficient bulk velocity to Doppler shift the line energy by up to
$\sim$200 eV.

\begin{figure}
\includegraphics[width=185mm]{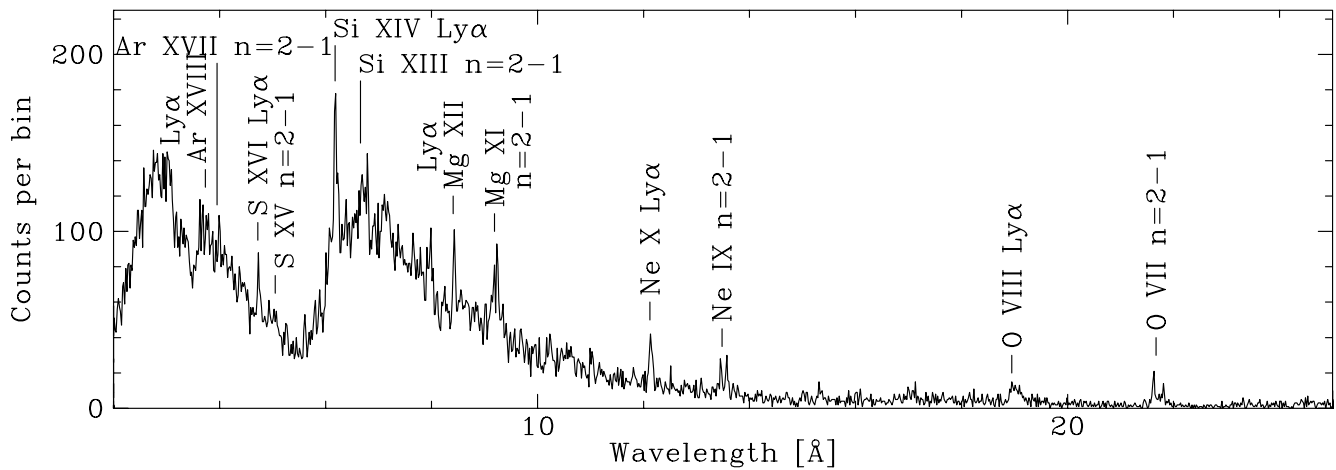}
\caption{The Chandra HETG (MEG) spectrum of V1223 Sgr, plotted against
wavelength.}
\end{figure}

The MEG data (Fig.6) reveal a rich emission-line spectrum, comparable to that
of EX Hya observed with the ASCA SIS.  Lines are detected from H- and He-like
ions of O, Ne, Mg, Si, S, and probably Ar.  Moreover, the MEG resolves
the resonant, intercombination, and forbidden triplets from He-like
ions of O, Ne, and Mg.  The forbidden lines are very weak, suggesting
relatively high density, and the intercombination lines are comparable
to the resonant components, suggesting photoionization origin.
Moreover, we do not detect strong Fe L features; their weakness is
another evidence for photoionization origin.
Finally, when plotted in 10 eV bins, the presence of the OVII
warm-absorber edge at 0.739 keV is strongly suggested.

It is therefore likely that these features originate in the immediate
pre-shock flow.  The only other body near the emission region, the
white dwarf itself, has a much higher density and therefore a lower
ionization parameter value than is required to produce these features.
As we have already speculated, the pre-shock
flow on the other hand is a natural origin both for the warm absorber
and of the photoionized plasma emission.  Detailed analyses are in
progress to quantify the conditions of the pre-shock flow using
these new diagnostics opened up by the Chandra grating instrument.

We are left with a major question, however: Where are the expected
thermal lines from the post-shock, cooling plasma?  
The presence of a strong Fe L bump in the ASCA spectrum of EX Hya,
among other clues, suggests that it is the lines from coronal plasma
that we have detected with the SIS.  Analyses of Chandra and XMM-Newton
grating observations of this system are expected to confirm this
interpretation.  Even though similar K shell features are seen,
V1223~Sgr appears completely different in this respect.  Comparison
of ASCA spectra suggest that V1223~Sgr is more typical of magnetic
CVs, whereas EX~Hya belongs to a smaller group perhaps with 3 other
systems, characterized perhaps by low accretion rates.

One possible explanation for the difference between the two magnetic
CVs (perhaps two groups of magnetic CVs) is the following.  In the
Aizu model, the plasma density increases sharply towards the bottom
of the shock.   Eventually, the post-shock region becomes
optically thick in the continuum.  In EX~Hya, this happens at about
0.65 keV, as suggested by the analysis of Fujimoto and Ishida (1997),
while in V1223~Sgr, this happens at a much higher temperature, say
kT$\sim$2 keV as has been suggested by Beardmore et al (2000).
Earlier works calculated the optical depth across the accretion
column just below the shock; their conclusion that that region
is optically thin remains valid.  However, most of the observed
medium energy (0.5--10 keV) X-rays actually originate much lower
down in the post-shock region (Fig.7).  For example, this plot
shows that, for a $kT_s$=20 keV shock, we need to use E$\sim$40 keV
continuum photons to probe the upper, low-density part of the
post-shock region.  Most of the photons we observe with ASCA,
for example, must come from the bottom $\sim$1\% of the shock!
The situation is more extreme for low energy line photons:
the emissivity of K$\alpha$ lines of elements up to Si and S,
as well as those of the Fe L lines, peaks at temperatures below 2 keV
in a collisionally ionized plasma.  In an Aizu shock with $kT_s$=20 keV,
this is the bottom $\sim$0.1\% of the shock, where the density is
$\sim$20 times greater than immediately below the shock.

These line photons cannot escape freely by travelling vertically, either,
partly because of the photoelectric absorption in the cool, pre-shock,
flow.  Moreover, the vertical electron scattering optical depth within
the post-shock region is non-negligible: 
the characteristic cooling time of the post-shock plasma is
inversely proportional to the local accretion rate, and is $\sim$1 s
at 1 g\,cm$^{-2}$s$^{-1}$.  Thus we expect $\sim$1 g\,cm$^{-2}$ of
vertical column in the post-shock region; in fact, a numerical integration
which properly accounts for the deceleration shows that the actual
number is $\sim$2 g\,cm$^{-2}$, with a weak dependence on the
primary mass.  It therefore appears that most of the low energy line
photons from the bottom can be hidden by a significant optical depth;
exactly how this works depends on the shock geometry.  In addition,
the white dwarf atmosphere has a scale height of several kilometers,
when heated to $kT\sim 20$ eV (this applies only to the area surrounding
the accretion column; underneath the column, the ram pressure will reduce
the scale height).  Thus, there may be a puffed up ring of white
dwarf atmosphere, further reducing our ability to see the low
energy line emissions.  Thermal Fe K$\alpha$ lines, which come
from higher up in the post-shock region, is probably responsible
for the 6.7 and 6.97 keV features seen in V1223~Sgr.  The shock
models would suggest that the Fe K$\alpha$ emitting region is optically
thin in the continuum.  This is consistent with the Hellier et al. (1998)
interpretation of the broad Fe K$\alpha$ lines, which requires a continuum
optical depth of 0.05--1.0 in the line emitting region, where the
temperature is kT$\sim$5 keV.

\begin{figure}
\includegraphics[width=110mm]{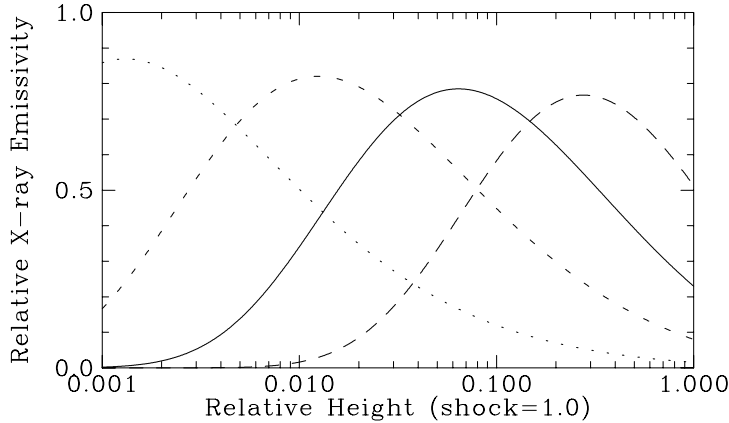}
\caption{The relative emissivity of E=2$kT_s$ (long dashes),
1.0$kT_s$ (solid), 0.5$kT_s$ (medium dashes), and 0.2$kT_s$
(short dashes) continuum photons as a function of shock height in
an Aizu model.}
\end{figure}

The accretion rate also controls the efficiency of pre-shock flow
heating.  Low accretion rate shocks are so tall that most of the
radiation escapes from the side, lowering the degree of photoionization
just above the shock.  Thus, EX~Hya could have very little emission
lines from this region while V1223~Sgr is dominated by it.

\section*{CONCLUSIONS SO FAR AND FUTURE PROSPECTS}

The Aizu shock model attempts to connect the optically thick region
immediately below the shock to the extremely optically thick white
dwarf atmosphere.  With hindsight, it is therefor natural that
optical depth plays a significant role in shaping the observed
X-ray spectra of magnetic CVs.  It appears that the low energy
X-ray line features of V1223~Sgr and EX~Hya have different origins,
despite some superficial similarities.  There is evidence that
the low energy lines seen in V1223~Sgr are from a dense,
photoionized plasma, probably in the pre-shock flow.  The lines
in EX~Hya is from the cooling post-shock plasma; these are weaker
or hidden in V1223~Sgr.  It is still likely that the Fe K$\alpha$
lines share the same origins (fluorescence off the white dwarf
surface for the 6.4 keV line, thermal plasma emission for the other
two components), since photoionization is unlikely to be strong
enough to produce a significant Fe K$\alpha$ feature in any CVs.
It has been suggested that resonant trapping of these line photons
is the cause for the broadening of these thermal Fe K$\alpha$ lines
in some systems; We await the recently approved Chandra grating
observation of AO~Psc to explore the details of this mechanism.
Along with more observations, a more detailed modelling is also
necessary.  However, if optical depth does play an important roll
in shaping the X-ray spectrum, as I have argued in this paper,
then a realistic geometry must be used in numerical simulations.
Therefore, it appears essential  to improve our knowledge of the
detailed shape of the post-shock region through, e.g., the observations
of eclipses and self-occultations of the X-ray emitting region.

What can we expect for non-magnetic CVs?  Despite the differences in
geometry, X-ray emitting regions in magnetic and non-magnetic systems
share some essential features: boundary layer emission in the latter
must originate in a shock-heated plasma which cools down, and becomes
denser, before it can settle down onto the primary surface.  The depth
of the gravitational potential, the size of the emitting region,
and the local accretion rate are all similar between magnetic and
non-magnetic CVs.  Therefore, it is possible that the optical
depth effects become important below some critical temperature,
hiding any intrinsic line features, in non-magnetic as well as in
magnetic systems.  With future high quality X-ray spectra, we may
begin to detect broadened lines, warm absorbers, and photoionized
plasma emissions in non-magnetic CVs, a very exciting prospect indeed.

\section*{ACKNOWLEDGEMENTS}

I thank Chris Done and Coel Hellier for their critical reading of this
manuscript.  I also thank Manabu Ishida, and Tim Kallman for the insights
that they have shared with me over the years.

\end{document}